\documentclass{nature}
\usepackage{graphicx}
\makeatletter
\let\saved@includegraphics\includegraphics
\AtBeginDocument{\let\includegraphics\saved@includegraphics}
\renewenvironment*{figure}{\@float{figure}}{\end@float}
\makeatother

\usepackage{multirow}
\usepackage{subfigure}
\usepackage{mathtools}
\usepackage[colorlinks=true,linkcolor=blue,citecolor=black,urlcolor=blue]{hyperref}
\usepackage{authblk}
\usepackage{lineno}
\usepackage[labelfont=bf]{caption}
\usepackage[figurename=Figure]{caption}
\usepackage{siunitx}
\usepackage[dvipsnames]{xcolor}
\usepackage{soul}
\usepackage{xfrac}

\begin{document}

\title{Defect-implantation for the all-electrical detection of non-collinear spin-textures}
\author[1*]{Imara Lima Fernandes}
\author{Mohammed Bouhassoune}
\author[1*]{Samir Lounis}
\affil{Peter Gr\"{u}nberg Institut and Institute for Advanced Simulation, Forschungszentrum J\"{u}lich and JARA, D-52425 J\"{u}lich, Germany}
\affil[*]{i.lima.fernandes@fz-juelich.de, s.lounis@fz-juelich.de}

\maketitle

\begin{abstract}
The viability of past, current and future devices for information technology hinges on their sensitivity to the presence of impurities. The latter can lead to resistivity anomalies, the so-called Kondo effect, reshape extrinsically Hall effects or reduce the efficiency of magnetoresistance effects essential in spintronics. Here we demonstrate that atomic defects ena\-ble highly efficient all-electrical detection of spin-swirling textures, in particular magnetic skyrmions, which are promising bits candidates in future spintronics devices. Impurities amplify the bare transport signal and  can alter significantly the spin-mixing magnetoresistance (XMR) depending on their chemical nature. Both effects are monitored in terms of the defect-enhanced XMR (DXMR) as shown for \textit{3d} and \textit{4d} transition metal defects implanted at the vicinity of skyrmions generated in PdFe bilayer deposited on Ir(111). The ineluctability of impurities in devices promotes the implementation of  DXMR in reading architectures with immediate implications in magnetic storage technologies. 
\end{abstract}

Defects are inherent to all devices and materials. Being unavoidable, they dramatically reshape transport properties, often negatively, and thus are key ingredients in settling the competitiveness of newly proposed technologies and therefore their survival. 
 In the context of spintronics~\cite{Wolf2001} exploiting the spin rather than the charge to carry information, defects can reduce the efficiency of magnetoresistance effects\cite{Evetts1998,Ke2010} in current perpendicular-to-plane  geometries such as  the giant magnetoresistance (GMR)\cite{Baibich1988,Binasch1989} or tunneling magnetoresistance (TMR)\cite{Julliere1975,Moodera1995}.  Impurities intrinsically give rise to inelastic transport channels allowing the exploration of electron-bosons interactions~\cite{Stipe1998,Heinrich2004,Schweflinghaus2014} while generating extrinsic contributions to Hall  effects~\cite{Nagaosa2010,Gradhand2010_PRL,Lowitzer2011,Fert2011,Zimmermann2016,Bouaziz2016}  in current-in-plane geometries. All of this is not surprising since the very fundamental Kondo effect\cite{Kondo1964} results from diluted magnetic impurities leading to an anomalous behaviour of the resistance at low temperature\cite{Daybell1968}.

Recently a new kind of magnetoresistance effect, the spin-mixing magnetoresistance (XMR), has been discovered~\cite{Crum2015,Hanneken2015}, which enables the all-electrical detection of non-collinear magnetic states, such as magnetic skyrmions and spin spirals. In contrast to the GMR or TMR effect, XMR requires a single magnetic electrode instead of two and thus its importance in establishing spin-swirling textures with chiral or topological protection properties, in particular skyrmions~\cite{Bogdanov1989,Roessler2006}, as future bits for information technology~\cite{Fert2017}. 
While several investigations were devoted to the impact of inhomogeneities on the motion and stability of skyrmions\cite{Fert2013,Sampaio2013,Iwasaki2013a,Iwasaki2013b,Jiang2015,Woo2016,Hanneken2016,Jiang2016,Litzius2017,LimaFernandes2018,Choi2017,Maccariello2018}, their implications in the electrical detection are yet to be determined.   Similarly to the GMR and TMR effects, it is often expected that defects would reduce the XMR efficiency.

\begin{figure}
 \centering
 \includegraphics[width=\columnwidth]{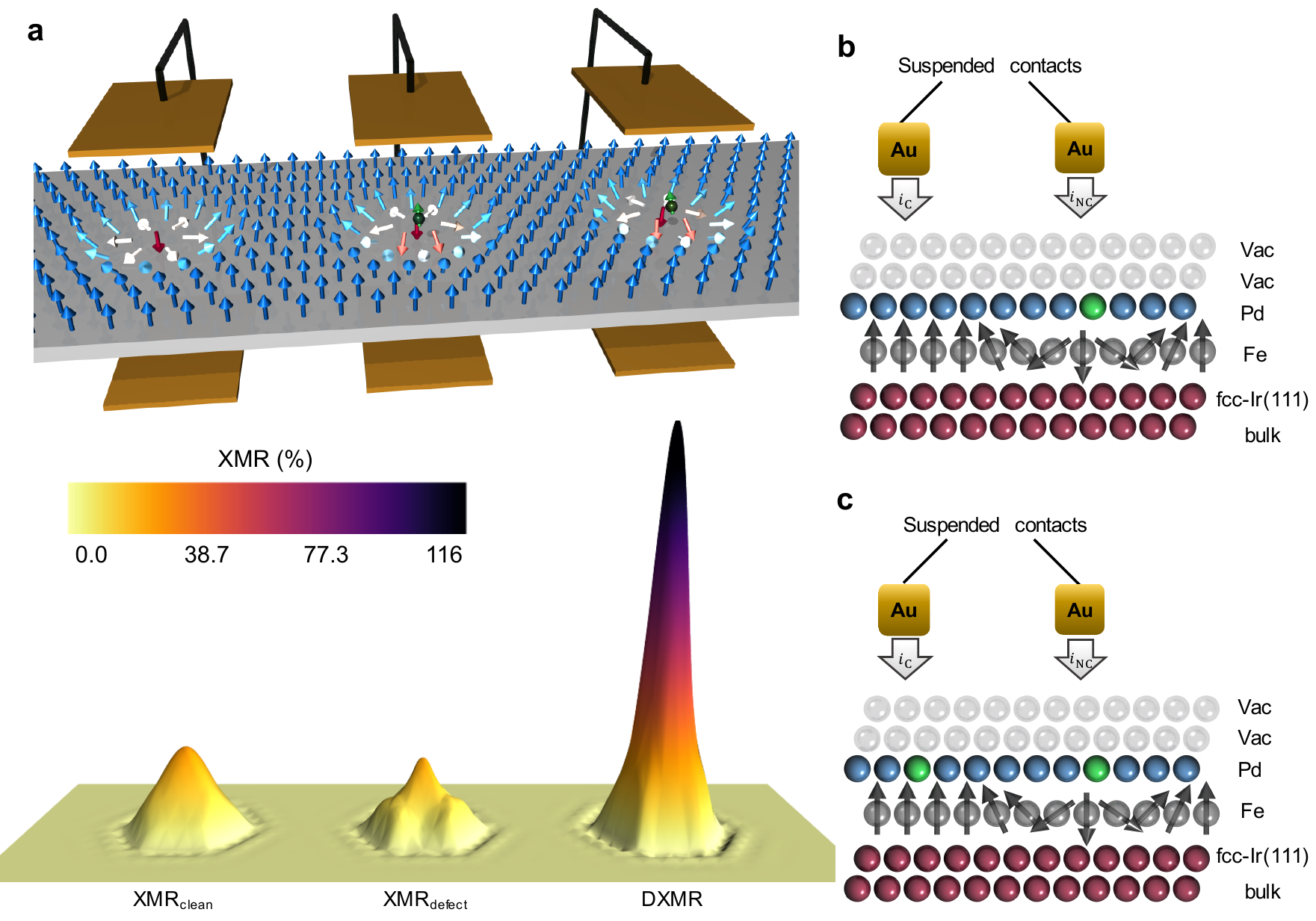}
 \caption{\textbf{All-electrical detection of magnetic skyrmions in the presence of defects.} \textbf{a}  Illustration of a tunneling transport experiment to read magnetic skymions on fcc-PdFe bilayer on the Ir(111) surface with the single-atomic defects. The spin-mixing of the electronic structure within the non-collinear environment leads to changes on the tunneling conductance allowing the magnetic data information to be sensed in a current perpendicular-to-plane geometry. The various XMRs signals, for the clean system and in the presence of defects, are shown on the the lower panel. Depending on the reference background two types of XMR signals can be measured, \textbf{b} while the clean ferromagnetic substrate leads to the novel defect-induced XMR (DXMR), \textbf{c} the substrate-defect complex background leads to the regular XMR$_{\text{defect}}$. The defects are indicated by green spheres. \label{Fig1}}
 \end{figure}

 In this article, we demonstrate from a full \textit{ab initio} approach that contrary to the current wisdom, atomic defects are of technological importance in reading non-collinear spin-states since they enable a highly efficient magnetoresistance signal, where the reference transport signal is amplified by many of the investigated impurities (see Figure~\ref{Fig1}a). 
 We envision scenarios, where impurities are manipulated atom-by-atom\cite{Eigler1990}, implanted\cite{Persaud2005} or spontaneously generated via intermixing mechanisms. 
 We introduce the defect-enhanced XMR (DXMR) in order to  evaluate the potential of defects in magnifying simultaneously the transport and XMR signals. We perform systematic simulations of atomic resolved transport measurements as probed within scanning tunneling microscopy/spectroscopy (STM/STS) of 3$d$ (V, Cr, Mn, Fe, Co, Ni) and 4$d$ (Nb, Mo, Tc, Ru, Rh) transition metal defects implanted in the Pd surface layer covering the fcc-Fe monolayer deposited on Ir(111) surface. The latter substrate is known to host few nanometers-wide magnetic skyrmions~\cite{Romming2013,Romming2015,Dupe2014} stabilized by the presence of Dzyaloshinskii-Moriya interaction~\cite{Dzyaloshinsky1958,Moriya1960}.  We identified the different mechanisms conspiring to make the universal trends of the various XMRs, as function of the impurities electronic states, distinct from each other. We put forward the DXMR effect as a key-reading tool in man-engineered defective substrates with immediate implications for device applications in the context of non-collinear states.

\section*{Results}

\subsection{Impact of defects on the all-electrical detection.} 

The all-electrical detection of single magnetic skyrmions via the XMR effect\cite{Crum2015,Hanneken2015,Kubetzka2017} is made possible via the non-collinearity of the Fe substrate's magnetic moments and the presence of spin-orbit interaction, which lead to a spin-mixing of the electronic states living initially in the majority- and minority-spin channels. Within tunneling transport experiments, these mixed states are detectable with a non-magnetic electrode (Figure 1a). In the case of PdFe/Ir(111) substrate, the XMR effect as probed in STM/STS is particularly enhanced for resonant  interface and surface states\cite{Bouhassoune2019}, which result from various hybridization mechanisms being affected by relativistic effects.

We consider a skyrmion passing at the vicinity of a defect as illustrated schematically in Fi\-gu\-re~\ref{Fig1}. Here, it is convenient to define two types of XMR signals contingent on the reference measurement with respect to which the efficiency of the MR effect is calculated. The reference point can be either the clean (defect-free) ferromagnetic substrate (see Figure~\ref{Fig1}b) or the substrate-defect complex (see Figure~\ref{Fig1}c) in the collinear magnetic ground state.
The former leads to the defect-enhanced XMR (DXMR) while the latter is a regular XMR signal but obtained in the pre\-sence of the defect and denoted in the following XMR$_{\mathrm{defect}}$. Such reference points can be conveniently measured in a device.
In both cases, we calculate the percent deviation of the conductance on top of a skyrmion from that of a reference point within a collinear magnetic region. We note that the XMR signal collected from an STM experiment  on a pinned skyrmion~\cite{Hanneken2015}  naturally includes the contribution of the pinning defect and should thus correspond to what we name DXMR.

Based on the Tersoff-Hammann approach\cite{Tersoff1983}, the differential conductance $\sfrac{dI}{dV}$ at a given bias voltage $V_\text{bias}$ measured  within STM/STS, equipped with a non-magnetic tip, is proportional to the local density of states (LDOS) of the substrate at the energy $\text{E} = \text{E}_\text{F} + eV_\text{bias}$ obtained  in  vacuum at the tip position. Here, we consider the tip located in second vacuum layer above the Pd substrate as indicated in Figure~\ref{Fig1}b-c, thus the efficiency of these magnetoresistance effects are calculated from:
\begin{eqnarray}
 \text{XMR}_{\text{defect}}\text{(E)} &=& \frac{\text{LDOS}^{\text{defect}}_{\text{NC}}\text{(E)}-\text{LDOS}^{\text{defect}}_{\text{C}}\text{(E)}}{\text{LDOS}^{\text{defect}}_{\text{C}}\text{(E)}},
\end{eqnarray} 
which is the equivalent of the conventional defect-free XMR, $\text{XMR}_{\text{clean}}$, where all LDOS are obtained in the clean region. C and NC correspond respectively to collinear and non-collinear magnetic regions, as shown in Figure \ref{Fig1}b-c.

The newly proposed DXMR efficiency is extracted from :
\begin{eqnarray}
\text{DXMR(E)} &=& \frac{\text{LDOS}^{\text{defect}}_{\text{NC}}\text{(E)}-\text{LDOS}^{\text{clean}}_{\text{C}}\text{(E)}}{\text{LDOS}^{\text{clean}}_{\text{C}}\text{(E)}}  \nonumber\\
&=&\text{XMR}_\text{clean}\text{(E)} + 
\frac{
\text{LDOS}^{\text{defect}}_{\text{NC}}\text{(E)}-
\text{LDOS}^{\text{clean}}_{\text{NC}}\text{(E)}}
{\text{LDOS}^{\text{clean}}_{\text{C}}\text{(E)}} \label{eq:dxmr_enhancement},\\ \nonumber
\end{eqnarray}
which obviously measures the enhancement of the XMR efficiency by the defect with respect to the signal obtained in the defect-free region. The latter equation shows that the DXMR signal is related to the ``traditional" effect and it is obviously shaped by how the tunneling matrix elements involving the defect are different from those of the defect-free substrate.  In fact, by re-expressing it as:
\begin{eqnarray}
\text{DXMR(E)} &=& 
\frac{
\text{LDOS}^{\text{defect}}_{\text{NC}}\text{(E)}-
\text{LDOS}^{\text{defect}}_{\text{C}}\text{(E)}}
{\text{LDOS}^{\text{clean}}_{\text{C}}\text{(E)}}+
\frac{
\text{LDOS}^{\text{defect}}_{\text{C}}\text{(E)}-
\text{LDOS}^{\text{clean}}_{\text{C}}\text{(E)}}
{\text{LDOS}^{\text{clean}}_{\text{C}}\text{(E)}},
\label{eq:dxmr2}
\end{eqnarray}
one deduces that besides the detection of the non-collinearity, the DXMR ratio measures the enhancement of the transport signal emanating from the reference points of the substrate because of the impurity. The larger it is, the better are the conditions for the all-electrical detection.

\begin{figure*}
\centering
\includegraphics[width=\textwidth,keepaspectratio]{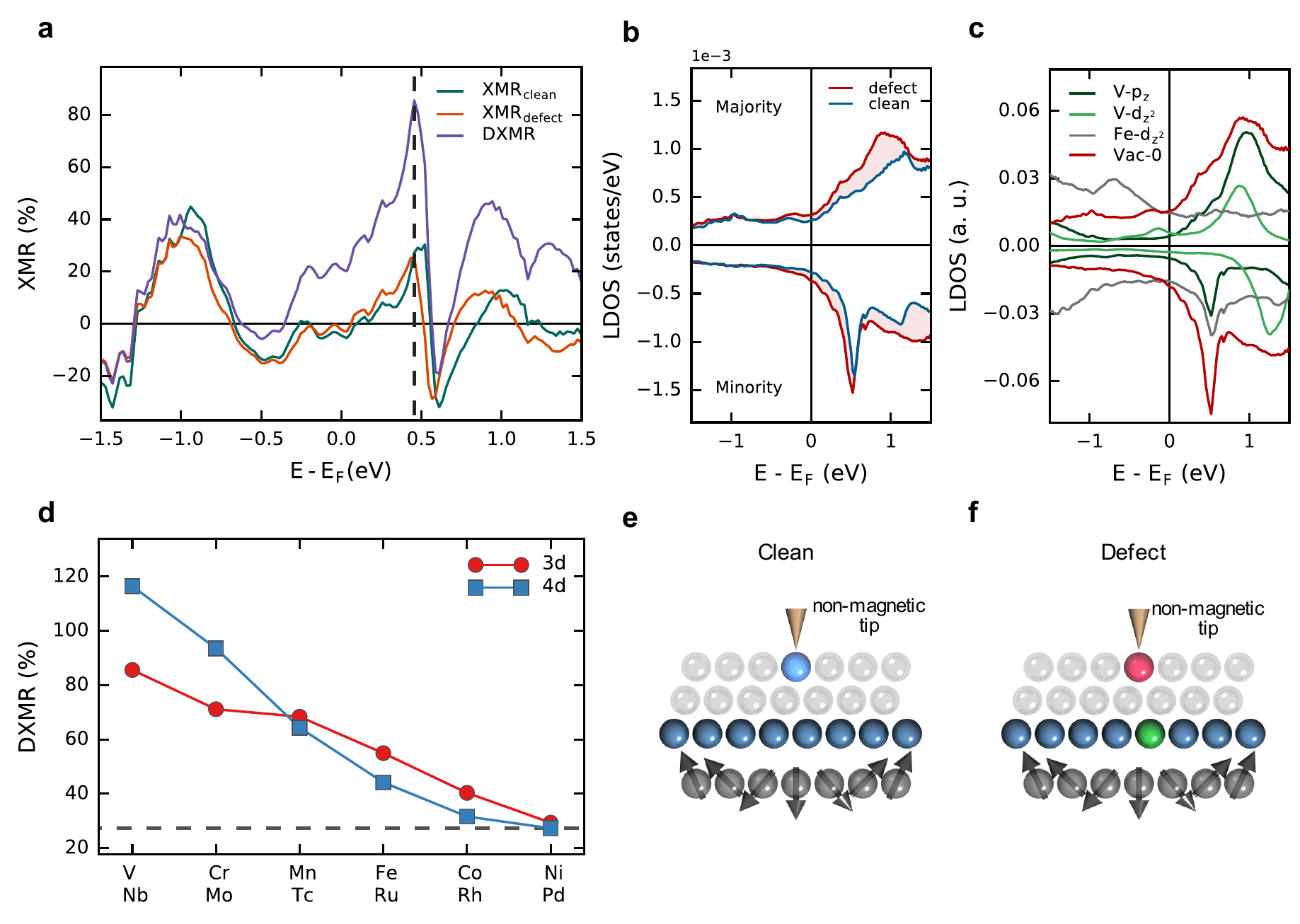}
\caption{\textbf{Efficiency of the XMR signal and the electronic structure.} \textbf{a} Comparison of various energy-resolved XMR signals measured on top of the skyrmion’s core considering the influence of a V impurity located close to skyrmion's core. The dashed line indicates the energy at which the DXMR reaches the maximum efficiency. \textbf{b} Electronic structure in the vacuum on top of the core of the skyrmion core resolved into majority and minority spin-channels for the clean system (blue line) and with a V-impurity (red line). \textbf{c} Superposition of the electronic states in arbitrary units of vacuum, V and Fe at the skyrmion's core. \textbf{d} Impact of chemical nature of the defect on the DXMR signal at an injection energy $eV_{\text{bias}} \approx \text{\SI{0.45}{\electronvolt}}$ for $3d$ (red circles) and  $4d$ (blue squares) impurities located close to the skyrmion center. The gray dashed line indicates the signal of the XMR$_{\text{clean}}$. \textbf{e-f} Illustrative legend of the STM-tip probing the clean (defective) substrate at the location defined by the light blue (red) sphere. The green sphere indicates the impurity. \label{fig:txmr_V-pure}}
\end{figure*}

For conciseness, we focus our first analyses on the impact of a V impurity at the closest vicinity of the skyrmion's core on the various XMRs, whose energy-resolved signals are plotted in Figure~\ref{fig:txmr_V-pure}a. The investigated skyrmion has a diameter $D_{Sk} \approx \SI{2.2}{\nano\meter}$ and we recall that in the skyrmion's core, the magnetic moment of the substrate Fe atom is flipped with respect to the ferromagnetic surrounding. Among the XMR signals, the DXMR efficiency is the largest with an impressive $\approx$ 85\%, i.e. an increase of $\approx$ 230\% with respect to the defect-free signal, for an energy injection about $eV_{\text{bias}} \approx \text{+\SI{0.45}{\electronvolt}}$ (Figure~\ref{fig:txmr_V-pure}a, purple curve). As deduced from Eq.~\ref{eq:dxmr2}, this large efficiency is just an indication that the defect enhances the bare tunneling transport signal. For comparison, the XMR$_{\text{clean}}$ (Figure~\ref{fig:txmr_V-pure}a, green curve) and XMR$_{\text{defect}}$ (Figure~\ref{fig:txmr_V-pure}a, orange curve) signals reach a similar efficiency around 26\%. In the whole investigated energy window, the latter two signals are very similar in magnitude and have a shape, which in general, seems similar to the DXMR signal. 

The XMR$_{\text{clean}}$ and XMR$_{\text{defect}}$ efficiencies are shaped by the same mechanisms, i.e. the non-collinearity of the substrate magnetic moments and the presence of spin-orbit interaction. In contrast, the DXMR signal is more sensitive to the hybridization of the electronic states of the defect with the substrate Fe atoms and the way these states decay into vacuum, which settles the tunneling matrix elements. In general, the enhancement factor of the DXMR signal is found similar to the one induced on the collinear substrate by implanted impurities. 

\subsection{Origin of enhancement of the bare transport signal and of the DXMR efficiency.}

The reference background for both the XMR$_{\text{clean}}$ and the DXMR signal is the defect-free region away from the skyrmion (collinear magnetic region). Thus, to understand the enhancement of the DXMR effect, we invoke Equation~\ref{eq:dxmr_enhancement} and analyse the vacuum LDOS just above the skyrmion core with and without the implanted V-atom (Figure~\ref{fig:txmr_V-pure}e-f). As illustrated in Figure~\ref{fig:txmr_V-pure}b, the vacuum electronic states of both spin-channels are strongly amplified by the presence of the V-defect affecting the differential conductance by virtue of the Tersoff-Hamann approximation. Here one identifies the main advantage of a substrate hosting impurities. The overall STM/STS signal is increased, which greatly facilitates the detection of buried skyrmions, as it is for PdFe/Ir(111) surface. Thus, defects can act as mediating probes for the electrical detection.
The observed vacuum states originate from the decay of the V and Fe substrate's electronic states plotted in Figure~\ref{fig:txmr_V-pure}e-f, as can be noticed from the one-to-one correspondence between the atomic virtual bound states and the features probed in vacuum. The orbital-resolved LDOS, e.g. the $p_{z}$ and $d_{z^2}$ states, leading to the largest tunneling matrix elements are shown in particular for the V defect and for the Fe atom at the skyrmion's core. The latter defines the spin-frame of reference in which the LDOS is plotted and the convention to denote the majority- and minority-spin states. 

Because of the large exchange splitting of V, the hybridization is weak between the occupied majority-spin states of Fe and the unoccupied ones of V, giving rise to the $d_{z^2}$ and $p_{z}$ virtual bound states located at $\approx\text{+ \SI{0.9}{\electronvolt}}$ and $\approx  \text{+ \SI{1.0}{\electronvolt}}$, respectively (see upper channel of Figure~\ref{fig:txmr_V-pure}c). Recalling that V replaces a Pd atom from the defect-free substrate, we expect thus a larger intensity of the features observed in vacuum because of the additional impurity-states. Similar conclusions can be drawn for the opposite spin channel. Interestingly,  
an unoccupied $p_{z}$ virtual bound state at $\approx \text{+ \SI{0.53}{\electronvolt}}$ is induced by the Fe minority-spin state  in the LDOS of V (see lower channel of Figure~\ref{fig:txmr_V-pure}c), with a larger intensity than the one obtained in the clean Pd-overlayer.

\subsection{Systematic trends.}
The enhancement of bare transport signal detected within the DXMR signal with respect to the conventional XMR is not limited to the V impurity but occurs for all investigated implanted defects of the $3d$ and $4d$ series. The factor of enhancement depends, however, on the chemical nature of the impurities as well as on the injection energy. In Figure~\ref{fig:txmr_V-pure}d, we systematically collect the efficiency of the DXMR as function of the impurities atomic number  for an injection energy of $eV_{\text{bias}} \approx \text{+ \SI{0.45}{\electronvolt}}$. Overall, the DXMR ratio decreases when increasing the atomic number of the defects. For the $3d$ elements, V leads to the highest efficiency, which is larger than the 30\% efficiency induced by Ni. The latter defect does not alter the defect-free XMR efficiency.  Interestingly, defects from the beginning of the $4d$ series lead to an impressive enhancement of the signal, reaching almost 116\% for Nb followed by 94\% for Mo which translate to a increase of about 350\% and 260\%, respectively, with respect to the defect-free XMR efficiency. This implies that the best impurities for the enhancement of the bare tunneling transport signal are early transition elements with a preference for the $4d$ series.

\begin{figure}[t!]
\centering
\includegraphics[width=\columnwidth,keepaspectratio]{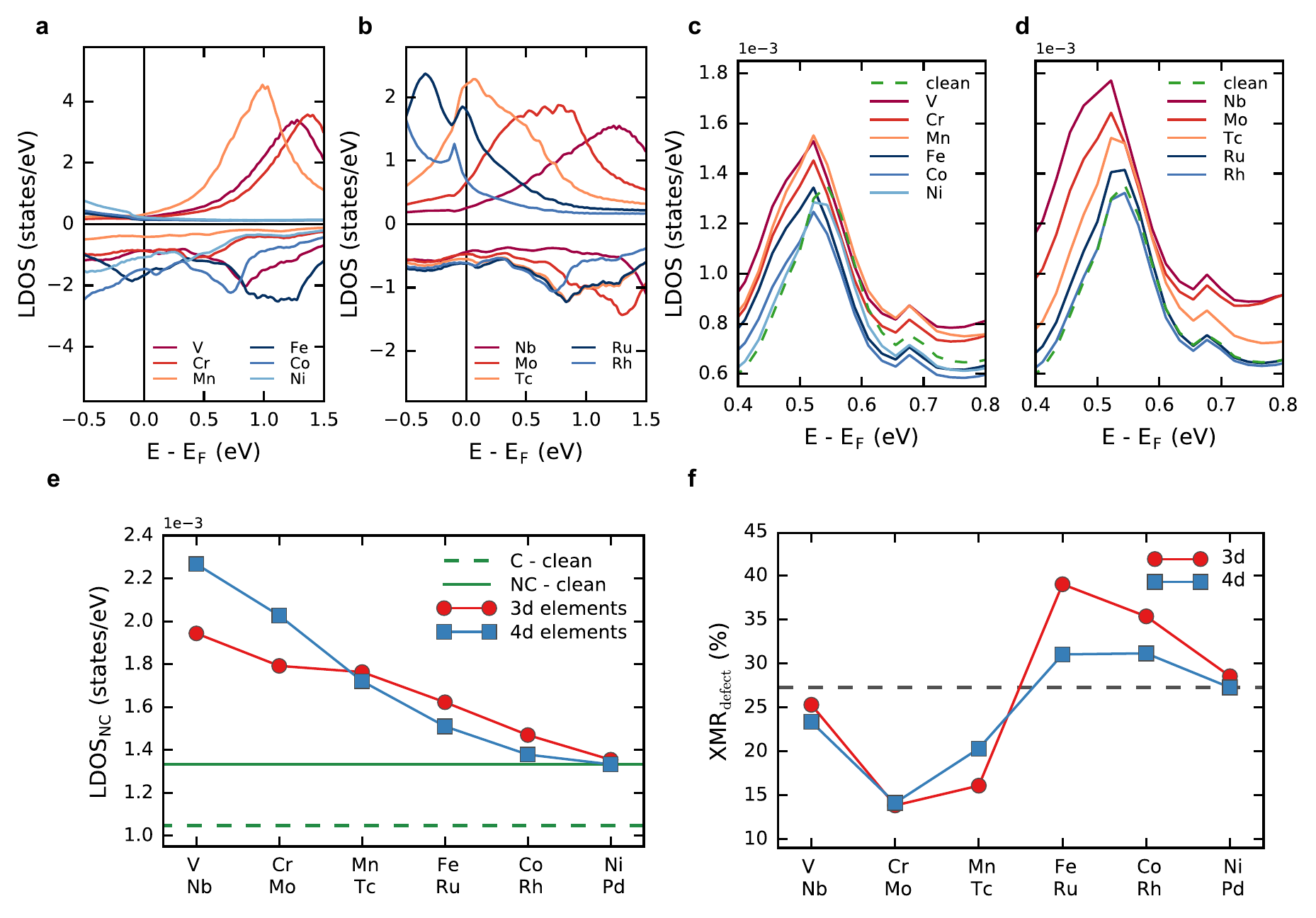}
\caption{\textbf{Impact of defects on the electronic structure and on the XMR$_{\text{defect}}$ signal.} Local density of states (LDOS) of the defect within the collinear configuration for \textbf{a} the $3d$-inatom series \textbf{b} the $4d$-inatom series. LDOS of the vacuum on top of the core of the skyrmion with \textbf{c} $3d$ and \textbf{d} $4d$ impurity close to the skyrmion core. \textbf{e} Chemical trend of the LDOS of the vacuum on top of the core of the skyrmion at $\text{+\SI{0.45}{\electronvolt}}$ for the $3d$-elements (red line) $4d$-elements (blue line). The green full line (dashed line) indicates the LDOS of the vacuum within the non-collinear (collinear) magnetic region for the defect free-system substrate at $\text{+\SI{0.45}{\electronvolt}}$. \textbf{f} Impact of band-filling on the defect spin-mixing magnetoresistance (XMR$_{\text{defect}}$) at an injection energy $eV_{\text{bias}} \approx \SI{0.45}{\electronvolt}$. The gray dashed line indicates the signal of the XMR$_{\text{clean}}$. }
    \label{Fig_dos3d-4d_inatom}
\end{figure}

The behavior of the DXMR signal as function of the chemical nature of the defect can be directly related to the filling of the impurities electronic states. By moving from left to right across the $3d$ ($4d$) atomic row of the periodic table the unoccupied states of the impurities shift towards the Fermi energy becoming partially or almost fully occupied at the end of the series as can be seen in  Figure~\ref{Fig_dos3d-4d_inatom}a(b) for the collinear magnetic states. Interestingly, this sequence is accompanied with a transition from an antiferromagnetic coupling to the substrate for V (Nb), Cr (Mo) and Mn (Tc) to a ferromagnetic coupling for Fe (Ru), Co (Rh) and Ni (Pd). The decrease of the impurities LDOS upon the filling of the electronic states,  as illustrated by Figure~\ref{Fig_dos3d-4d_inatom}e, explains the DXMR trend plotted in Figure~\ref{fig:txmr_V-pure}d as well as the very large ratio induced by the $4d$ impurities when compared to the $3d$ ones. The smaller exchange splitting of the $4d$ elements increases the possibility of having large LDOS around the Fermi energy.

Paying closer attention to the electronic features of the vacuum on top of the core of the skyrmion  responsible for the discussed effect for the $3d$ and $4d$ series (see Figure~\ref{Fig_dos3d-4d_inatom}c-d), one identifies that the peaks obtained for the antiferromagnetic V (Nb), Cr (Mo) and Mn (Tc) are more intense than those of the ferromagnetic Fe (Ru), Co (Rh) and Ni (Pd). Here, two concomitant mechanisms are at play. First the hybridization strength is known to decrease from left-to-right and to increase from up-to-down across the transition element series of the peridioc table~\cite{LimaFernandes2018}. Second, the $p_z$ state induced in the electronic structure of the impurities depends on the magnetic coupling because of the switch of the spin-nature of the impurity states hybridizing with the minority-spin states of Fe. Thus, more impurity states are available when the spin-alignment is rather ferromagnetic.

\subsection{Real-space XMRs contrast and inhomogeneous non-collinearity.}
Contrasting the DXMR behavior, the $\text{XMR}_{\text{defect}}$ for an injection energy $eV_{\text{bias}} \approx \text{
+ \SI{0.45}{\electronvolt}}$ when plotted as function of the atomic number (see Figure~\ref{Fig_dos3d-4d_inatom}f) exhibits a S-like shape with a maximum and a minimum close to the middle of the series with the efficiency increasing up to about 56\% with respect to the defect-free signal. 
By changing the injection energy, the S-shape can be reversed and the location of the extrema can be strongly shifted. Interestingly, while the bare transport signal is enhanced for the early transition elements, the XMR signal is better magnified by impurities with more than half-filled $d$-states.

\begin{figure*}
\centering
\includegraphics{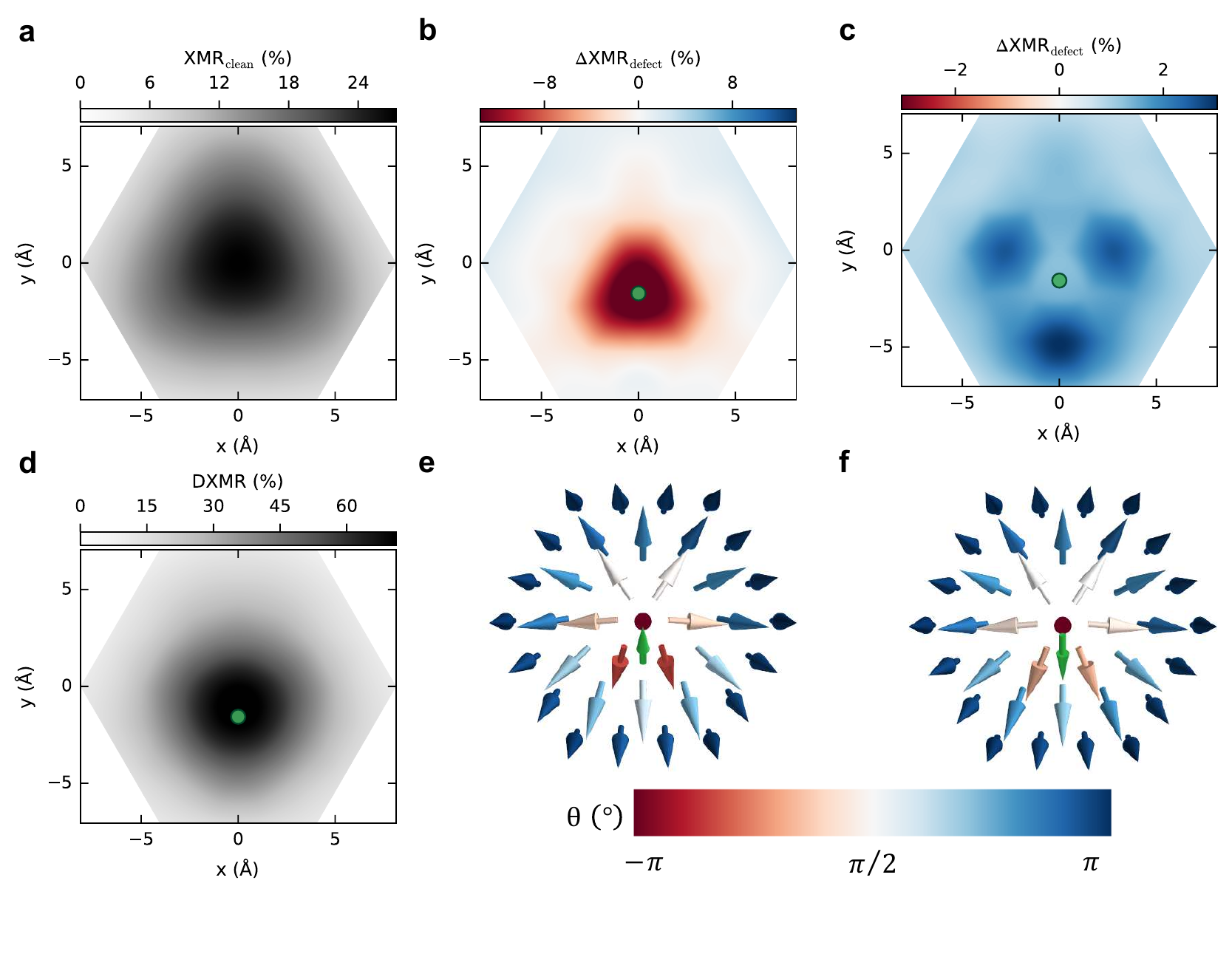}
\caption{\textbf{Impact of single atomic defects on XMR-signals of skyrmions.} \textbf{a} Defect-free XMR signal compared to \textbf{d} DXMR signal induced by a Cr impurity. Difference between XMR$_\text{defect}$ and XMR$_\text{clean}$ plotted for \textbf{b} Cr and \textbf{c} Ni impurities close to the core of skyrmion. The green circle indicates the position of the impurity. The corresponding skyrmion magnetic textures are shown in \textbf{e} for Cr and \textbf{f} for Ni, where the green arrow represents the impurity magnetic moment while the rest of arrows are those of the Fe substrate atoms. The angle $\theta$ is the polar angle defining the magnitude of the z-component of the magnetization for each substrate atom. \label{Fig5}}
\end{figure*}

The XMR efficiency can be of use to detect slight changes in the magnetic texture. The XMR$_{\text{clean}}$ and XMR$_{\text{defect}}$ depend on the opening angle between neighboring magnetic moments. In the defect-free region, the measurable electrical contrast is expected to exhibit a highly symmetric shape,  translating the symmetry of the magnetic texture, with the highest intensity at the center of the skyrmion (see Figure~\ref{Fig5}a). 
At the vicinity of a defect, the skyrmion experiences an asymmetric environment impacting its spin-texture and consequently the all-electrical XMR contrast. This is better grasped by visualizing the difference $\Delta \text{XMR}_{\text{defect}} = \text{XMR}_{\text{defect}} - \text{XMR}_{\text{clean}}$ as done in  Figure~\ref{Fig5}b-c for Cr and Ni impurities. For Cr, the $\text{XMR}_{\text{defect}}$ signal strongly decreases at the close vicinity of the impurity since the strong exchange coupling of the impurity and the substrate atoms reduces the surrounding non-collinearity (see Figure~\ref{Fig5}e), which lowers the XMR ratio. For Ni, however, the change in the XMR signal (Fi\-gu\-re~\ref{Fig5}c) is off-set from the defect with the pattern being less intense than for Cr, which corresponds to the modifications induced in the magnetic texture shown in Figure~\ref{Fig5}f.

The DXMR signal is not that sensitive to the impurity-induced non-collinear modifications and reaches a maximum value on top of all defects. 
As an example, the Cr-related DXMR signal at a bias energy $\approx \text{\SI{0.45}{\electronvolt}}$ is illustrated in  Figure~\ref{Fig5}d when the impurity is close to the core of the skyrmion.

\section*{Discussion}

In this work, we explore the possibility of using defects to detect non-collinear magnetic textures using all-electrical means. By strongly modifying the tunneling matrix elements, defects increase the intensity of the background transport signal. This would prove useful in case of buried non-collinear spin-textures, which are difficult to access with surface probe techniques. To track the impact of impurities, we defined the DXMR efficiency, which monitors the enhancement of the bare transport signal and of the XMR efficiency induced by the presence of the defects. From the obtained general trends upon filling of the impurities electronic states, our investigations point to the use of early transition elements to enhance the tunneling transport patterns, while elements with more than half-filled $d$-shells seem to be better suited for a better XMR-detection.

Contrary to the usual expectations, defects provide a path to enhance in various ways the all-electrical detection of non-collinear magnetic textures. The DXMR and XMR$_{\text{defect}}$ can be incorporated in future reading technologies, which enforces the view that controlled engineering of defective-materials is a promising route for device architectures.

\begin{methods}

\subsection{Computational details.}

The calculations were performed with the full-potential  scalar-relativistic Korringa-Kohn-Rostoker (KKR) Green function method with spin-orbit coupling\cite{Papanikolaou2002,Bauer2013}. The defect and the single magnetic skyrmions are considered using the embedded technique which allows us to evaluate details of the electronic structure at different positions. The Pd/Fe/Ir(111) slab consists of a fcc-stacked PdFe bilayer deposited on 34 layers of Ir. The embedded cluster consists of 37 Fe atoms and 124 atoms in total, details of the self-consistent calculations can be found at Ref.\cite{LimaFernandes2018}. The local density of states for the ferromagnetic-state or skyrmion-state were obtained performing one single interaction with a k-mesh of $200 \times 200$ and an angular momentum cutoff of $l_{\text{max}} = 3$ for the orbital expansion of the Green function.

\subsection{Data availability} The data that support the findings of this study are available from the corresponding authors on request.

\end{methods}

\begin{addendum}

\item This work is supported by the European Research Council (ERC) under the European Union's Horizon 2020 research and innovation programme (ERC-consolidator grant 681405 — DYNASORE). We acknowledge the computing time granted by the JARA-HPC Vergabegremium and VSR commission on the supercomputer JURECA at Forschungszentrum Jülich.

\item[Author contributions] I.L.F. performed the first-principles calculations. All authors discussed the results and helped writing the manuscript.

\item[Competing Interests] The authors declare that they have no competing financial interests. 

\item[Correspondence] Correspondence and requests for materials should be addressed to I.L.F. (email: i.lima.fernandes@fz-juelich.de) or to S.L. (email: s.lounis@fz-juelich.de).

\end{addendum}

\maketitle

\section*{References}
\bibliographystyle{naturemag}

\end{document}